\documentclass[11pt,twoside]{article}

\usepackage{asp2006}
\usepackage{graphicx}

\begin{document}

\title{The Chemical Evolution of LSB Galaxies}   

\author{Lars Mattsson,  
Brady Caldwell, and 
Nils Bergvall}  

\affil{Dept. of Astronomy \& Space Physics, Uppsala University, Sweden}  
 
\begin{abstract} 
  We have derived oxygen and nitrogen abundances of a sample of late-type, low surface brightness (LSB) galaxies 
  found in the Sloan Digital Sky Survey (SDSS). Furthermore, we have computed a large grid (5000 models) of chemical 
  evolution models (CEMs) testing various time-scales for infall, baryon densities and several power-law initial 
  mass functions (IMFs) as well. Because of the rather stable N/O-trends found both in CEMs (for a given IMF) and in 
  observations, we find that the hypotheses that LSB galaxies have stellar populations dominated by low-mass stars, 
  i.e., very bottom-heavy IMFs (see Lee et al. 2004), can be ruled out. Such models predict much too high N/O-ratios 
  and generally too low O/H-ratios. We also conclude that LSB galaxies probably have the same ages as their high 
  surface brightness counterparts, although the global rate of star formation must be considerably lower in these 
  galaxies. 
\end{abstract}

\section{Late-type LSBGs in the SDSS}
We use the data compilation by Caldwell \& Bergvall (2006) consisting of a sample of 1199 close to edge-on ($b/a\leq 0.25$) low surface 
brightness (LSB) galaxies in the fourth data release of the Sloan Digital Sky Survey (SDSS-DR4). All galaxies in the sample
were selected in order to obtain a sample of isolated, bulge-less systems, which is ideal for comparison with simple CEMs since the
merger histories of such galaxies have less influence on the evolution of elemental ratios. This total sample is divided 
by $g-r$ colour limits into a blue Sample A (377 galaxies), a green (intermediate) Sample B (436 galaxies) and a red Sample C 
(386 galaxies). From these 1199 galaxies we pick out the ones with a relative error in the H$\alpha$-flux less than $25\%$.

\section{Numerical Model Grid}   
  We have computed a large grid (5000 models) of one-zone CEMs, testing time-scales for infall from $\tau_{\rm inf} = 1.0$ Gyr 
  to $\tau_{\rm inf} = 7.9$ Gyr, baryon densities ranging between $\Sigma_{\rm final} = 1.0 M_\odot$ pc$^{-2}$ and 
  $\Sigma_{\rm final} = 6000.0 M_\odot$ pc$^{-2}$, which should cover most of the possible variations within dwarf and late-type galaxies. 
  Star formation is prescribed by a simple Schmidt-law, $\dot{\Sigma}_\star = \eta\,\Sigma_{\rm gas}^{1.5}$, 
  where the constant $\eta$ is varied between $\eta = 0.001$ to $\eta = 0.2$. Furthermore, 
  we have tried several power-law initial IMFs $\phi(m)\sim m^{-(1+x)}$ in order to test the bottom-heavy
  IMF hypothesis ($x=2.85$) suggested by Lee et al. (2004). The nucleosynthesis prescriptions are taken from Chieffi \& Limongi (2004)
  for high mass stars and van den Hoek \& Groenewegen (1997) for low and intermediate stars and all model tracks where evolved
  over 13.7 Gyr (one Hubble-age).

  \begin{figure}[!ht]
  \includegraphics[width=4.33cm]{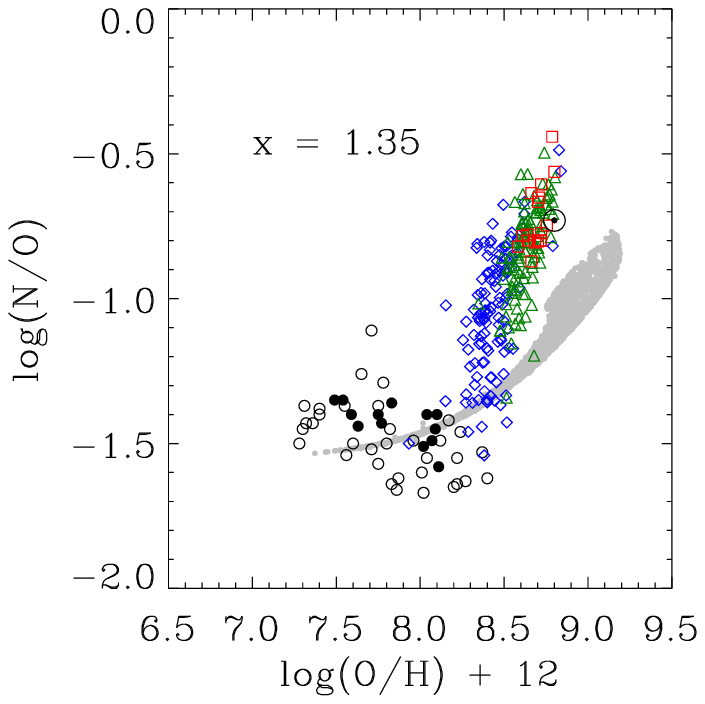}
  \includegraphics[width=4.33cm]{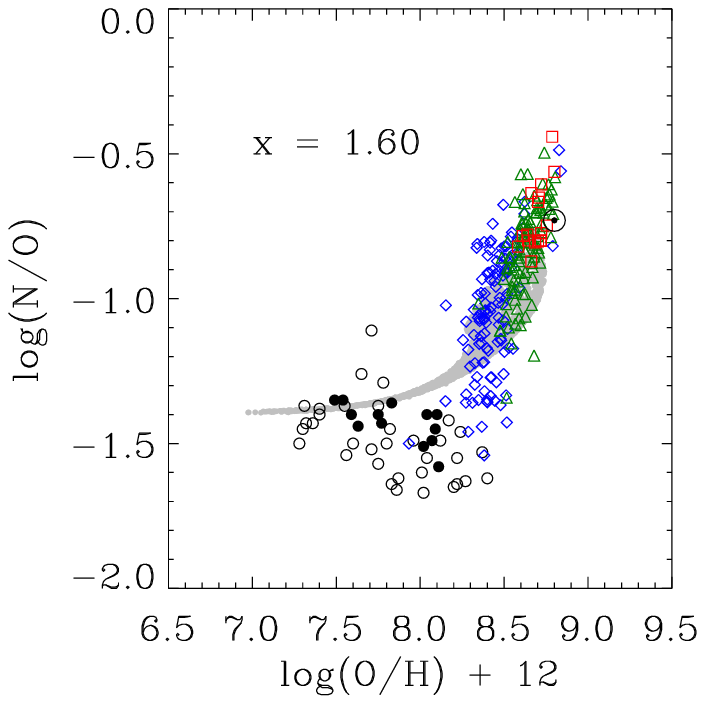}
  \includegraphics[width=4.33cm]{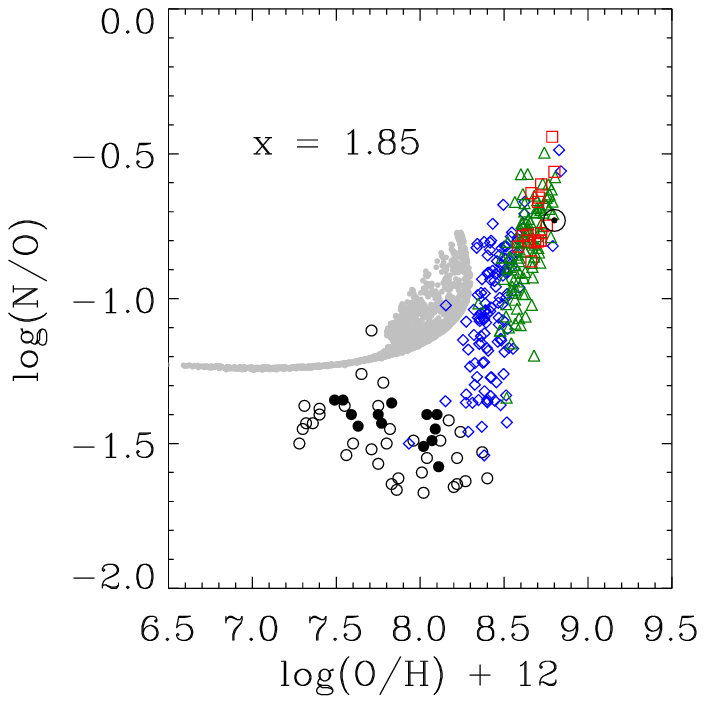}
  \caption{  \label{models} Observed abundances and model results (grey dots) for different IMFs. Slowly evolving
  tracks populate the $\log(\rm{N/O})$-plateau, while rapidly evolving tracks populate the "secondary branch". 
  The abundances of the late-type LSBGs are shown as blue diamonds (Sample A), green triangles (Sample B) and red
  squares (Sample C). Black dots and circles show data for dwarf LSBGs from R\"onnback \& Bergvall (1995), van Zee 
  et al. (1996) and van Zee \& Haynes (2006).
  }
  \end{figure}

\section{Results and Conclusions}
In Fig. 1 we show a comparison between our abundances derived from spectral data from SDSS-DR4 using the so-called P-method
(see, e.g., Pilyugin 2003) and our model results. Clearly, the late-type LSBGs do not deviate from the general, rising trend 
found in HSB galaxies. Based on the results from our model grid we draw the following conclusions:\\[3mm]
{
- The $\log({\rm N/O})$-ratios in dwarf irregular and late-type LSBGs follow the same basic pattern as their HSB counterparts.\\[1mm]
- The chemical evolution, as revealed by the HII-regions, is not compatible with an extremely bottom-heavy IMF as suggested
      by Lee et al (2004), nor is it compatible with a top-heavy IMF.\\[1mm]
- The LSB property of these objects is most likely due to a low global star formation density, although the efficiency of
      star formation in star-forming regions must be as high as in HSBGs in order to explain the high $\log(\rm{N/O})$-ratios.\\[1mm]
- Given these results, it is unlikely (although not impossible) that the ages of LSBGs are significantly different from the 
      ages of corresponding HSBGs. 
}

\end{document}